%% file: main.tex
\documentclass[5p,times]{elsarticle}

\usepackage{amsmath}
\usepackage[ruled,linesnumbered,lined,noend]{algorithm2e}
\usepackage{algorithmic} 
\usepackage{graphicx}
\usepackage{textcomp}
\usepackage{xcolor}
\usepackage{comment}
\usepackage{breakurl}
\usepackage{xspace}
\usepackage{pifont}
\usepackage{multirow}
\usepackage{amsfonts}
\usepackage{listings}
\usepackage{subfigure}
\usepackage{color}
\usepackage{url}
\usepackage{booktabs}
\usepackage{colortbl}
\usepackage{tcolorbox}
\usepackage{hyperref}
\usepackage{ulem}
\usepackage{fvextra}
\usepackage{enumitem}

\definecolor{light-gray}{gray}{0.85}

\definecolor{verylightgray}{rgb}{.97,.97,.97}

\newcommand{\tool}{\textsc{CoTR}\xspace}
\newcommand{\toolAttack}{\textsc{CoTR-A}\xspace}
\newcommand{\toolDefense}{\textsc{CoTR-D}\xspace}

\journal{Information and Software Technology}

\begin{document}


\begin{frontmatter}
	
\title{Assessing and Improving Syntactic Adversarial Robustness of Pre-trained Models for Code Translation}

\author[NUAA]{Guang Yang}
\ead{yang.guang@nuaa.edu.cn}

\author[NUAA]{Yu Zhou\corref{mycorrespondingauthor}}
\cortext[mycorrespondingauthor]{Corresponding author}
\ead{zhouyu@nuaa.edu.cn}

\author[NUAA]{Xiangyu Zhang}
\ead{zhangx1angyu@nuaa.edu.cn}

\author[NTU]{Xiang Chen}
\ead{xchencs@ntu.edu.cn}

\author[Birkbeck]{Tingting Han}
\ead{t.han@bbk.ac.uk}

\author[Birkbeck]{Taolue Chen\corref{mycorrespondingauthor}}
\ead{taolue.chen@gmail.com}

\address[NUAA]{Nanjing University of Aeronautics and Astronautics, Nanjing, China}
\address[NTU]{School of Information Science and Technology, Nantong University, Nantong, China}
\address[Birkbeck]{Birkbeck, University of London}

\begin{abstract}

\textbf{Context:} Pre-trained models (PTMs) have demonstrated significant potential in automatic code translation. However, the vulnerability of these models in translation tasks, particularly in terms of syntax, has not been extensively investigated.

\textbf{Objective:} To fill this gap, our study aims to propose a novel approach {\tool} to assess and improve the syntactic adversarial robustness of PTMs in code translation.

\textbf{Method:} {\tool} consists of two components: {\toolAttack} and {\toolDefense}. {\toolAttack} generates adversarial examples by transforming programs, while {\toolDefense} proposes a semantic distance-based sampling data augmentation method and adversarial training method to improve the model's robustness and generalization capabilities.
The Pass@1 metric is used by {\tool} to assess the performance of PTMs, which is more suitable for code translation tasks and offers a more precise evaluation in real-world scenarios. 

\textbf{Results:} The effectiveness of {\tool} is evaluated through experiments on real-world Java$\leftrightarrow$Python datasets. The results demonstrate that {\toolAttack} can significantly reduce the performance of existing PTMs, while {\toolDefense} effectively improves the robustness of PTMs.

\textbf{Conclusion:} Our study identifies the limitations of current PTMs, including large language models, in code translation tasks. It highlights the potential of {\tool} as an effective solution to enhance the robustness of PTMs for code translation tasks.


\end{abstract}


\end{frontmatter}

\input{sections/1introduction}
\input{sections/2preliminaries}
\input{sections/3approach}
\input{sections/4experiment}
\input{sections/5result}

\input{sections/6discussion}
\input{sections/7related}
\input{sections/8conclusion}

\section*{Acknowledgement}
This work was partially supported by the National Natural Science Foundation of China (NSFC, No.\ 62372232), the Postgraduate Research \& Practice Innovation Program of Jiangsu Province (No.\ KYCX23\_0396), and the Collaborative Innovation Center of Novel Software Technology and Industrialization. 
T.\ Chen is partially supported by an oversea grant from the State Key Laboratory of Novel Software Technology, Nanjing University (KFKT2022A03) and Birkbeck BEI School Project (EFFECT).

\section*{Declaration of Competing Interests}
The authors declare that they have no known competing financial interests or personal relationships that could have appeared to influence the work reported in this paper.
	
\section*{CRediT Authorship Contribution Statement}
\textbf{Guang Yang:} Data curation, Software, Writing - original draft.
\textbf{Yu Zhou:} Conceptualization, Methodology, Writing -review \& editing, Supervision.
\textbf{Xiangyu Zhang:} Data curation, Software, Validation.
\textbf{Xiang Chen:} Writing - review \& editing, Validation.
\textbf{Tingting Han:} Writing - review \& editing, Validation.
\textbf{Taolue Chen:} Writing - review \& editing, Validation.

\normalem
\bibliography{main}
\bibliographystyle{elsarticle}

\vspace{1cm}
\noindent\textbf{Guang Yang} received the M.D. degree in computer technology from Nantong University, Nantong, in 2022. Then he is currently pursuing the Ph.D degree at Nanjing University of Aeronautics and Astronautics, Nanjing.
His research interest is AI4SE and he has authored or co-authored more than 20 papers in refereed journals or conferences, such as ACM Transactions on Software Engineering and Methodology, Empirical Software Engineering, Journal of Systems and Software, International Conference on Software Maintenance and Evolution (ICSME), and International Conference on Software Analysis, Evolution and Reengineering (SANER).
More information about him can be found at:

\url{https://ntdxyg.github.io/}
\par
\vspace{1cm}

\vspace{1cm}
\noindent\textbf{Yu Zhou} is a full professor in the College of Computer Science and Technology at Nanjing University of Aeronautics and Astronautics (NUAA). He received his BSc degree in 2004 and PhD degree in 2009, both in Computer Science from Nanjing University China. Before joining NUAA in 2011, he conducted PostDoc research on software engineering at Politechnico di Milano, Italy. From 2015-2016, he visited the SEAL lab at University of Zurich Switzerland, where he is also an adjunct researcher. His current research interests mainly generative models for software engineering, software evolution analysis, mining software repositories, and reliability analysis. He has been supported by several national research programs in China. More information about him can be found at: 

\url{https://csyuzhou.github.io/}.
\par
\vspace{1cm}

\vspace{1cm}
\noindent\textbf{Xiangyu Zhang} is currently pursuing a Master's degree at the College of Computer Science and Technology of Nanjing University of Aeronautics and Astronautics. His research interests include code generation and model interpretability.
\par
\vspace{1cm}

\vspace{1cm}
\noindent\textbf{Xiang Chen} received the B.Sc. degree in the school of management from Xi'an Jiaotong University, China in 2002. Then he received his M.Sc., and Ph.D. degrees in computer software and theory from Nanjing University, China in 2008 and 2011 respectively. He is currently an Associate Professor at the Department of Information Science and Technology, Nantong University, Nantong, China. He has authored or co-authored more than 120 papers in refereed journals or conferences, such as IEEE Transactions on Software Engineering, ACM Transactions on Software Engineering and Methodology, Empirical Software Engineering, Information and Software Technology, Journal of Systems and Software, Journal of Software: Evolution and Process, Automated Software Engineering, Journal of Computer Science and Technology, International Conference on Software Engineering (ICSE), The ACM Joint European Software Engineering Conference and Symposium on the Foundations of Software Engineering (ESEC/FSE), International Conference Automated Software Engineering (ASE), International Conference on Software Maintenance and Evolution (ICSME), International Conference on Program Comprehension (ICPC), and International Conference on Software Analysis, Evolution and Reengineering (SANER). His research interests include software engineering, in particular software testing and maintenance, software repository mining, and empirical software engineering. He received two ACM SIGSOFT distinguished paper awards in ICSE 2021 and ICPC 2023. He is the editorial board member of Information and Software Technology. More information about him can be found at: 

\url{https://smartse.github.io/index.html}
\par
\vspace{1cm}

\vspace{1cm}
\noindent\textbf{Taolue Chen} received the Bachelor and Master degrees from Nanjing University, China, both in computer science. He was a junior researcher (OiO) at the Centrum Wiskunde \& Informatica (CWI) and acquired the PhD degree from the Vrije Universiteit Amsterdam, The Netherlands. He is currently a lecturer at the School of Computing and Mathematical Sciences, Birkbeck, University of London. He had been a postdoctoral researcher at University of Oxford (UK)  and University of Twente (NL). His research area is software engineering with an emphasis on program analysis and verification. His present research focus is on the border of software engineering and machine learning. He applies verification and programming language techniques to improve the trustworthiness of machine learning models. Meanwhile, he applies data-driven approaches to support software development. He has published over 130 papers in journals and conferences such as POPL, LICS, CAV, ICSE, FSE, ASE, ETAPS (TACAS, FoSSaCS, ESOP, FASE), OOPSLA, NeurIPS, ICLR and IEEE TSE, ACM TOSEM, ACM TOCL. He won the Best Paper Award of SETTA’20 and the 1st Prize in the CCF Software Prototype Competition. He has served  editorial board or program committee for various international journals and conferences. More information about him can be found at 

\url{https://chentaolue.github.io/}
\par
\vspace{1cm}

\end{document}

%% file: sections/1introduction.tex
\section{Introduction}

Automated code translation is vital for seamless interoperability between systems and platforms during software migration~\cite{weisz2021perfection, weisz2022better}. 
It becomes particularly crucial when adopting new programming languages or modernizing legacy systems. However, manual code translation is time-consuming and error-prone~\cite{roziere2020unsupervised}. For example, the migration of COBOL to Java at the Commonwealth Bank of Australia took about five years and \$750 million to complete.
To address this challenge, researchers have developed automated code translation tools, which have recently demonstrated great potential through the adoption of pre-trained models (PTMs)~\cite{roziereleveraging, liu2023syntax}.


Despite the significant progress made in the field of PTMs, there are concerns regarding their accuracy and robustness in real-world scenarios. The previous studies~\cite{zhang2020adversarial, zhang2020generating, rabin2021generalizability, zeng2022extensive, zhou2022adversarial, yang2022natural, yang2022important} primarily focused on tasks, such as code summarization and method name prediction. 
In contrast, our study specifically focuses on the code translation task, which presents new challenges and needs dedicated research. During programming, there exist diverse alternatives to accomplish the same functionality. For example, one can use interchangeable for-loop or while-loop; or some developers may prefer \verb|a<b| but others may prefer \verb|b>a| when writing conditional statements. In the context of code translation, it is crucial to ensure these syntactically distinct yet semantically equivalent code snippets are translated into semantically equivalent target code. That means a \textit{robust} (neural) code translation model should recognize the semantics while not overfitting the syntax of source code. However, in reality, even the most sophisticated tools (such as Copilot~\cite{copilot}) have faced criticism regarding their robustness~\cite{vaithilingam2022expectation, Grover}. 
For example, a minor syntactic difference can completely alter the behavior of a program, leading to serious bugs. Our goal is to improve the robustness of existing PTMs in code translation. This would enable software engineering researchers and practitioners to understand the strengths and weaknesses of PTM-based code translators. Furthermore, this would provide insights and guidelines for developing more robust models. To this end, we utilize adversarial attack techniques and generate adversarial examples, exposing vulnerabilities and weaknesses that may not be apparent in real-world scenarios or existing evaluation methods.

\begin{figure}[t]
	\centering
	\subfigure[Syntactic errors in translation caused by loop exchange operation]{%
		\includegraphics[width=0.42\textwidth]{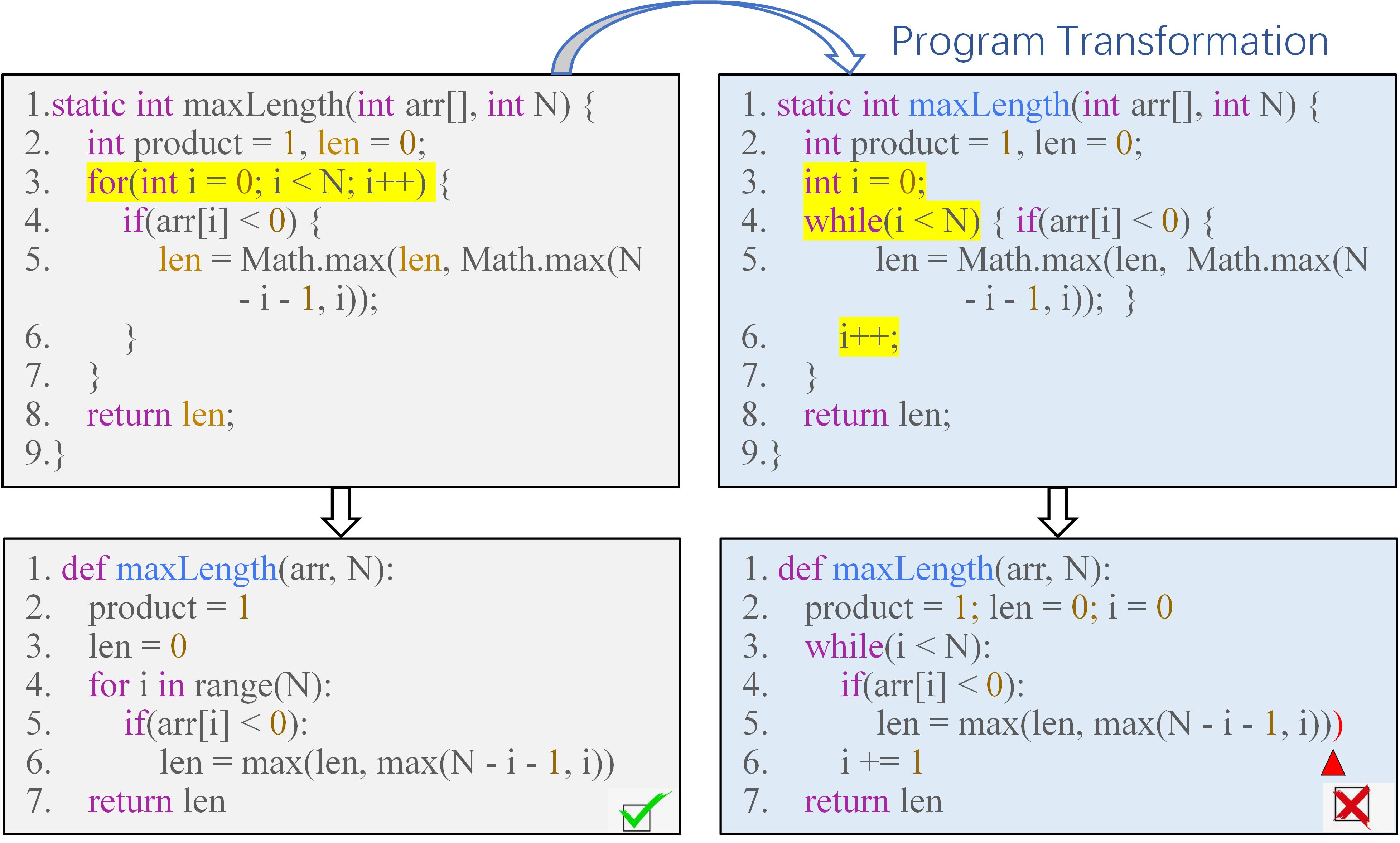}%
	}%
	\hfill
	\subfigure[Functional errors in translation caused by condition exchange operation]{%
		\includegraphics[width=0.42\textwidth]{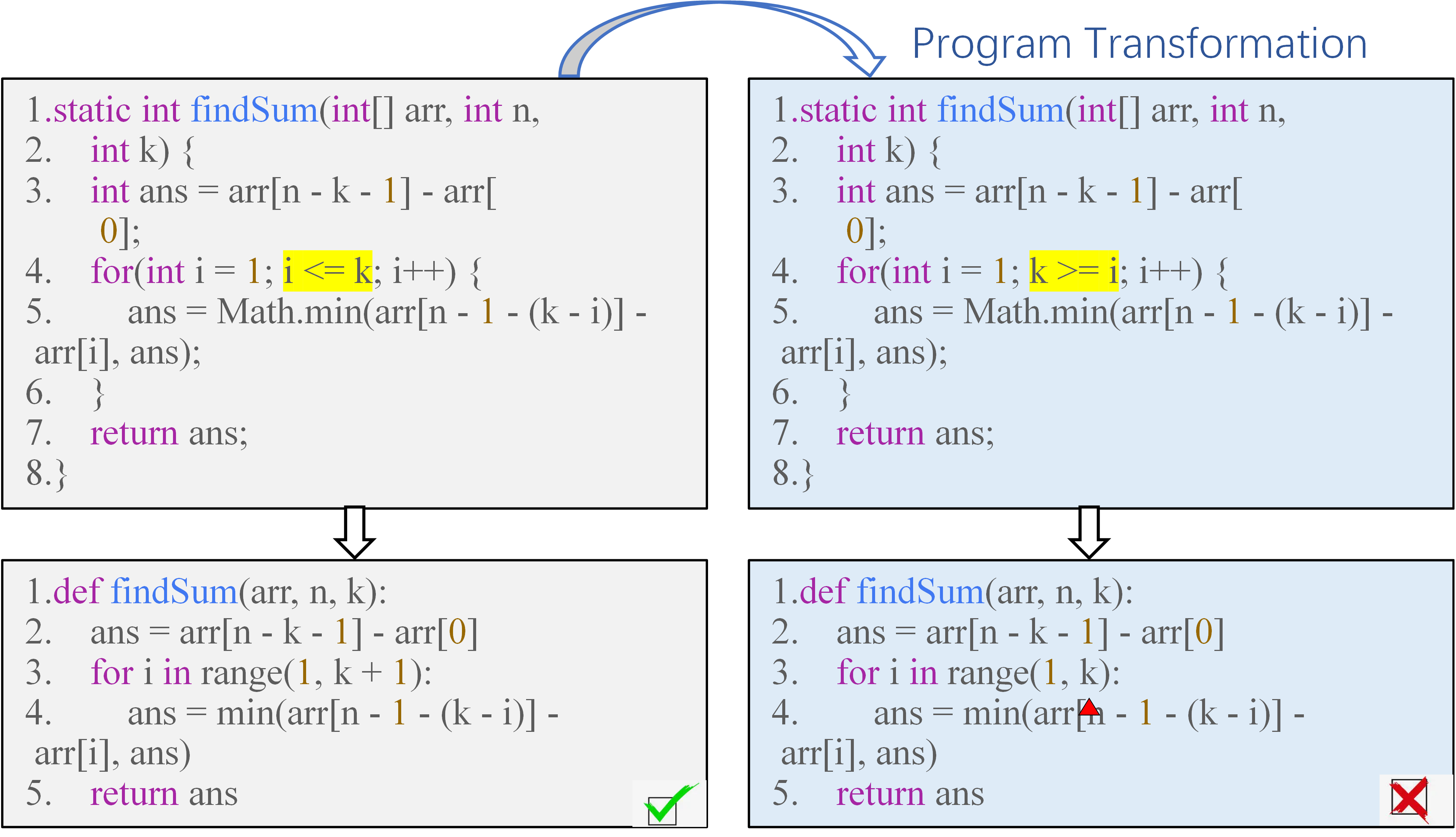}%
	}%
	\caption{Syntax and Functional Errors in Translated Code by CodeT5}
	\vspace{-0.4cm}
	\label{fig:exam}
\end{figure}

To illustrate the limitations of PTMs in code translation, we present two examples translated by CodeT5~\cite{wang2021codet5} in Figure~\ref{fig:exam} (from Java to Python). 

In the initial example, the source code undergoes the ``For/While Exchange" transformation (i.e., exchanging the while-loop for a for-loop, highlighted in yellow), leading to the detection of syntax errors in the CodeT5-translated code.
In the second example, the source code undergoes `Condition Exchange' transformation (highlighted in yellow), resulting in the emergence of functional faults in the CodeT5-translated code.
These faults highlight the vulnerability of PTMs in handling subtle (syntactic) changes in code.
%
%
Therefore, we provide a comprehensive investigation of the robustness of PTMs in code translation in this study.




In our study, we propose a novel approach {\tool} (\underline{Co}de \underline{T}ranslation Model \underline{R}obustness Detector), comprising two essential components: {\toolAttack} and {\toolDefense}.
{\toolAttack} 
imitates different programming styles through program transformation, attempting  
to generate code snippets to fail the model. 
In our study, this is referred to as an adversarial attack.  
Specifically, {\toolAttack} first defines a set of program transformation rules that are used to generate a collection of semantically equivalent source code. 
{\toolAttack} then feeds these code snippets into the victim model to identify the code that makes the model fail (i.e., does not pass all the test cases) as an adversarial code snippet. 
The low robustness of the model implies its sensitivity to the syntax of the input code, casting doubts about whether these seemingly well-performing models have truly learned essential code semantic features.

Different from AI security research, it is important to emphasize that the objective of {\toolAttack} is not to attack but to enhance the PTMs' performance. Consequently, {\toolDefense} adopts a dual-pronged strategy by retraining the victim model.
Firstly, {\toolDefense} augments the training data using program transformation techniques. To mitigate the risk of overfitting, {\toolDefense} computes the semantic distance between the original data and the augmented data, selecting the sample with the maximum distance for sampling. 
Although this approach effectively enhances the model's robustness, it may lead to a reduction in accuracy for specific models. To tackle this concern, {\toolDefense} additionally adopts a gradient-based adversarial training method. Through this dual strategy, {\toolDefense} achieves noteworthy improvements in model robustness without compromising performance.

We conducted a large-scale empirical study involving 10 state-of-the-art PTMs on a real-world dataset. Our investigation reveals that Encoder-Decoder PTMs outperform other PTMs in terms of performance.
Regarding robustness, our study unfortunately indicates that the existing PTMs are not sufficiently robust when it comes to code translation. Specifically, our {\toolAttack} reduces the Pass@1 metric by at least 17.97\% (from 17.97\% to 43.02\%) in the Java-to-Python dataset and by at least 14.29\% (from 14.29\% to 47.46\%) in the Python-to-Java dataset. 
Furthermore, we observed that the existing pre-training techniques for model robustness (e.g., contrast learning~\cite{liu2023contrabert} and adaptation learning~\cite{lu1codexglue}) are more adept at defending against token-based attacks~\cite{yang2022natural, yang2022important} but are less sensitive to the syntactic transformations proposed in this study.
Our findings also highlight the advantages of utilizing both data augmentation and adversarial training to enhance the robustness and generalization of code translation models.

We believe these findings are valuable for researchers and practitioners engaged in the field of code translation. For researchers, they provide valuable insights into the limitations and challenges faced by PTMs in code translation. This understanding can guide future research endeavors in developing more effective and reliable PTMs tailored specifically for code translation tasks. For practitioners, our study offers a practical solution to address the limitations of existing PTMs in code translation. By adopting our proposed tool, practitioners can enhance the accuracy of code translation, resulting in improved software quality. 

The contributions of our study are summarized as follows.
\begin{itemize}
    \item We construct high-quality datasets and comprehensively evaluate the functional accuracy and robustness of PTMs in code translation.
    \item We propose {\toolAttack}, which can effectively perform adversarial attacks on PTMs through program transformation. 
    \item We propose {\toolDefense}, a defense method that can achieve significant improvements in model robustness without sacrificing its performance.
\end{itemize}

To facilitate the reproducibility of our study, we release source code, benchmarks, and experimental data at \url{https://github.com/NTDXYG/COTR}. 

%% file: sections/2preliminaries.tex
\begin{table*}[h]
 \caption{Comparison of program transformation rules}
\centering

\resizebox{1.0\textwidth}{!}{
\begin{tabular}{c|l|l|l|ccc}
  \toprule
\textbf{Type} & 
\textbf{Method} & \textbf{Description} & \textbf{Example} & \textbf{S} & \textbf{I} & \textbf{R}\\
 \midrule
\multirow{6}{*}{
\begin{tabular}[c]{@{}l@{}}Token\\Renaming\end{tabular}
}
& API Renaming\cite{pour2021search, dongmixcode} & rename an API by other API names & \Verb|np.add()|$\rightarrow$ \Verb|np.sinc()| & $\times$ & $\times$ & $\times$  \\
 \cline{2-7}
& \begin{tabular}[c]{@{}l@{}}Arguments Renaming  \cite{pour2021search, zhou2022adversarial, jha2023codeattack, henke2022semantic, dongmixcode, zhang2020generating} \end{tabular} & rename an augment by other words & \Verb|def f(size)|$\rightarrow$ \Verb|def f(a)| & $\checkmark$ & $\times$ & $\times$  \\
 \cline{2-7}
& \begin{tabular}[c]{@{}l@{}}Local Variable Renaming \\ \cite{chakraborty2022natgen, zhang2020generating, pour2021search, wei2022cocofuzzing, jha2023codeattack, chen2022generating, rabin2022programtransformer, rabin2021generalizability, wang2022recode, zhou2022adversarial, henke2022semantic, zhang2022towards, jia2023clawsat, dongmixcode}\end{tabular} & \begin{tabular}[c]{@{}l@{}} rename a local variable by other words \\ and recursively update all related variables\end{tabular} & \Verb|number=1|$\rightarrow$ \Verb|size=1| & $\checkmark$ & $\times$ & $\times$  \\
 \cline{2-7}
& \begin{tabular}[c]{@{}l@{}}Method Name Renaming \\\cite{zhang2020generating, pour2021search, wei2022cocofuzzing, jha2023codeattack, wang2022recode, zhou2022adversarial, zhang2022towards, dongmixcode, yang2022important}\end{tabular} & rename a method by other words & \Verb|def count(a)|$\rightarrow$ \Verb|def f(a)| & $\checkmark$ & $\times$ & $\times$  \\
\midrule
\multirow{17}{*}{
\begin{tabular}[c]{@{}l@{}}Statement\\Insert
\end{tabular}
}
& Arguments Adding\cite{pour2021search, dongmixcode} & add an unused argument to function definition. & \Verb|def f(a)|$\rightarrow$ \Verb|def f(a, b)| & $\times$ & $\times$ & $\checkmark$  \\
 \cline{2-7}
& \begin{tabular}[c]{@{}l@{}}Dead Code Adding \\ \cite{chakraborty2022natgen, pour2021search, wei2022cocofuzzing, rabin2022programtransformer, rabin2021generalizability, wang2022recode, henke2022semantic, zhang2022towards, dongmixcode}\end{tabular} & \begin{tabular}[c]{@{}l@{}} add an unreachable or unused code at a \\ randomly selected location 
\end{tabular}& add: \Verb|if (1==0): print(0)| & $\checkmark$ & $\times$ & $\times$  \\
 \cline{2-7}
& \begin{tabular}[c]{@{}l@{}}Duplication Code Adding  \\ \cite{tian2021generating, wei2022cocofuzzing, dongmixcode}\end{tabular} & \begin{tabular}[c]{@{}l@{}} duplicate a randomly selected assignment\\ and insert it to its next line \end{tabular} & \Verb|a=1;|$\rightarrow$ \Verb|a=1;a=1;| & $\checkmark$ & $\checkmark$ & $\times$  \\
 \cline{2-7}
& \begin{tabular}[c]{@{}l@{}}Filed Enhancement  Adding  \cite{dongmixcode}\end{tabular} & \begin{tabular}[c]{@{}l@{}} enhance the rigor of the code by checking \\if the input of each argument is None \end{tabular} & \begin{tabular}[c]{@{}l@{}}\Verb|def f(a):|$\rightarrow$ add: \Verb|if a is|\\ \Verb|None: print("ERROR")| \end{tabular}& $\checkmark$ & $\times$ & $\checkmark$  \\
 \cline{2-7}
& \begin{tabular}[c]{@{}l@{}}Plus Zero  Adding  \cite{wei2022cocofuzzing, dongmixcode}\end{tabular} & \begin{tabular}[c]{@{}l@{}} select an numerical assignment of mathematical \\ calculation and plus zero to its value \end{tabular} & \Verb|a=1|$\rightarrow$ \Verb|a=1+0| & $\checkmark$ & $\checkmark$ & $\times$  \\
 \cline{2-7}
& \begin{tabular}[c]{@{}l@{}}Print Adding\\ \cite{pour2021search, rabin2022programtransformer, henke2022semantic, jia2023clawsat, dongmixcode} \end{tabular}& add a print line at a randomly selected location & add: \Verb|print(1)| & $\checkmark$ & $\times$ & $\times$  \\
 \cline{2-7}
& \begin{tabular}[c]{@{}l@{}}Return Optimal  Adding \\ \cite{pour2021search, dongmixcode}\end{tabular} & \begin{tabular}[c]{@{}l@{}} change the return content to a variant with the\\ same effect \end{tabular} & \begin{tabular}[c]{@{}l@{}}\Verb|return 1|$\rightarrow$ \Verb|return 0 if| \\  \Verb|(1==0) else 1| \end{tabular}& $\checkmark$ & $\times$ & $\times$  \\
 \cline{2-7}
& TryCatch Adding\cite{rabin2022programtransformer, henke2022semantic} & add a single  \Verb|try{A}catch(B){C} | statement & add: \Verb|try: catch():| & $\checkmark$ & $\times$ & $\checkmark$  \\
 \cline{2-7}
& UnrollWhiles Adding\cite{henke2022semantic} & \begin{tabular}[c]{@{}l@{}} add a randomly selected, while loop in the target \\ program has its loop body unrolled exactly one step \end{tabular} & \begin{tabular}[c]{@{}l@{}}\Verb|while(A){B}|$\rightarrow$ \\ \Verb|while(A){B;while(A){B} break;}| \end{tabular}& $\checkmark$ & $\times$ & $\times$  \\
\midrule
\multirow{10}{*}{
\begin{tabular}[c]{@{}l@{}}Statement\\Exchange\end{tabular}
}
& \begin{tabular}[c]{@{}l@{}} Loop Exchange \\ \cite{chakraborty2022natgen, pour2021search, tian2021generating, chen2022generating, rabin2022programtransformer, rabin2021generalizability, wang2022recode}\end{tabular} & \begin{tabular}[c]{@{}l@{}} replace a for loop with an equivalent while loop or\\ replace a while loop with an equivalent for loop\end{tabular} & \Verb|For|$\Leftrightarrow$ \Verb|While| & $\checkmark$   & $\checkmark$ & $\checkmark$  \\
 \cline{2-7}
& \begin{tabular}[c]{@{}l@{}} Expression Exchange \\ \cite{chakraborty2022natgen, tian2021generating, chen2022generating}\end{tabular} & use the properties of expressions to transform & \Verb|a+=b|$\rightarrow$ \Verb|a=a+b| & $\checkmark$ & $\checkmark$ & $\checkmark$  \\
 \cline{2-7}
& \begin{tabular}[c]{@{}l@{}} Permute Exchange \\ \cite{chakraborty2022natgen, rabin2022programtransformer, rabin2021generalizability} \end{tabular} & swap two independent statements in a basic block & \begin{tabular}[c]{@{}l@{}} \Verb|if(a){A} else{B}| $\rightarrow$ \\ \Verb|if(!a){B} else{A}| \end{tabular} & $\checkmark$ & $\checkmark$ & $\checkmark$  \\
 \cline{2-7}
& \begin{tabular}[c]{@{}l@{}} Condition Exchange \\ \cite{pour2021search, rabin2021generalizability, chakraborty2022natgen, chen2022generating, rabin2022programtransformer, wang2022recode} \end{tabular} & \begin{tabular}[c]{@{}l@{}} reorder the left and right parts of a binary condition or\\transform \Verb|True| and \Verb|False| by logical operations \end{tabular}& \begin{tabular}[c]{@{}l@{}} \Verb|if(a>b)|$\rightarrow$ \Verb|if(b<a)| \\ or \Verb|True|$\rightarrow$ \Verb|!False| \end{tabular} & $\checkmark$ & $\checkmark$ & $\checkmark$  \\
 \cline{2-7}
& \begin{tabular}[c]{@{}l@{}} Switch/If Exchange \\ \cite{tian2021generating, rabin2022programtransformer, rabin2021generalizability} \end{tabular}& replace a switch statement with a if-else statement & \Verb|Switch|$\Leftrightarrow$ \Verb|If/Else| & $\checkmark$ & $\checkmark$ & $\checkmark$  \\
  \bottomrule
\end{tabular}}
 \label{tab:rules}
\vspace{-0.4cm}
\end{table*}

\section{Preliminaries}
\label{sec:preliminaries}


\subsection{Code Translation}
Code translation models take source code snippets as input and generate corresponding code snippets in the target language. In general, the model is trained on a labeled dataset $\mathcal{D}_{train}=\left(\mathcal{X}, \mathcal{Y}\right): =\{\left(x_1, y_1\right), \cdots,\left(x_N, y_N\right)\}$, where each $x_i \in \mathcal{X}$ (resp. $y_i \in \mathcal{Y}$) represents a source (resp. target) code snippet.
Most pre-trained code translation models utilize the Transformer~\cite{vaswani2017attention} architecture. The model $\mathcal{M}$, which comprises an encoder and a decoder, accepts the source code snippet $x \in \mathcal{X}$ as input and produces a sequence of hidden states $\mathcal{H}\left(x\right)={h_1(x), h_2(x), \dots, h_n(x)}$ as encoder's output. The decoder then accepts the hidden states as well as the previously generated target code token $y_{1:t-1}$ as input to generate the probability distribution over the next target token $y_t$. This is achieved by passing the last decoder hidden state $\mathbf{s}_t$ through a linear layer followed by a softmax activation function 
\[ 
P_{\Theta_M}\left(y_t \mid y_{1: t-1}, x\right)=\operatorname{softmax}\left(\mathbf{W} \mathbf{s}_t+\mathbf{b}\right) ,
\]
where $\mathbf{W}$ and $\mathbf{b}$ are the learnable parameters of the linear layer.  The negative log-likelihood is usually used as the loss function
 \[  
\mathcal{L}\left(x, y; \Theta_M\right)=-\sum_{t=1}^T \log P_{\Theta_M}\left(y_t \mid y_{1: t-1}, x\right) ,
\]
where  $T$ denotes the length of the target code sequence and $\Theta_M$ denotes the set of parameters of $\mathcal{M}$. During the training process, $\mathcal{M}$ is optimized to minimize the negative log-likelihood of the target code sequence presented given the source code sequence over the labeled data sampled from $\mathcal{D}_{train}$, i.e.,
\[ 
 \min_{\Theta_M} \mathbb{E}_{(x, y) \sim \mathcal{D}_{train}}\left[ \mathcal{L}\left(x, y ; \Theta_M\right)\right] 
\]

Please note that not all PTMs adopt the encoder-decoder structure. For instance, GPT-like PTMs solely consist of decoders, making the step where the encoder obtains hidden states optional.


\subsection{Program Transformation}
\label{subsec:program}

Program transformation is a technique that modifies source code without compromising its overall functionality~\cite{mens2004survey}, and it has found extensive application in software engineering. The process of program transformation begins by parsing the code into an abstract syntax tree. Subsequently, depending on the transformation rule, the appropriate node is identified, and the transformation is executed accordingly.
%
%
In general, program transformation can be formalized as a function $\mathcal{F}$ that takes the source code $x$ and a set of transformation rules $\mathcal{R}$ as inputs and produces a set $T$ of transformed code that satisfies the given constraints $\mathcal{G}$.
%
$T = \mathcal{F}(x, \mathcal{R}, \mathcal{G})$.

To conduct a systematic review of the existing literature on program transformation, we employed a rigorous methodology. Firstly, we identified relevant keywords and conducted a comprehensive search for papers. We then manually screened the titles and abstracts of the papers to eliminate irrelevant ones. Additionally, we utilized academic search engines to supplement our search by checking citation status and exploring the list of published papers from relevant researchers.
Finally, we have curated a list of program transformation rules from the literature (until March 2023), as presented in Table~\ref{tab:rules}. These rules are categorized into three groups: `Token Renaming', `Statement Insert', and `Statement Exchange'. These rules can be analyzed from three distinct aspects: semantics (S), informativeness (I), and readability (R). Semantics refers to whether the transformed code preserves the same functionality as the original code. Informativeness~\cite{yuan2021bartscore} pertains to whether the transformed code is consistent with the intended information expressed in the original code. Readability assesses whether the transformed code aligns with the human readability of the original code.
It is important to note that in the literature, the concept of semantics can be understood from at least two different perspectives: one in the sense of formal semantics, capturing the functionality of the code, while the other is often referred to as ``naturalness"~\cite{HindleBSGD12}, which treats the code as text in a natural language. In this study, we use semantics and informativeness to refer to these two perspectives of semantics, respectively.

Among these three types, we assert that only rules under the type of `Statement Exchange' can maintain functional consistency, informativeness, and readability (refer to the 5th column of Table~\ref{tab:rules}). On the other hand, the remaining two types of program transformation are highly likely to impact at least one of these three aspects.
As an illustrative example, let us consider the Method Name Renaming rule. This rule involves modifying the method name in the code, but it may result in a loss of information. Specifically, the method name often contains valuable information about the functionality of the code from a natural language perspective. Replacing it with a generic name such as  `f' could lead to a loss of such essential information.


%% file: sections/3approach.tex
\section{The {\tool} approach}
\label{sec:method}

\begin{figure*}[h]
\centering
\includegraphics[width=0.85\textwidth]{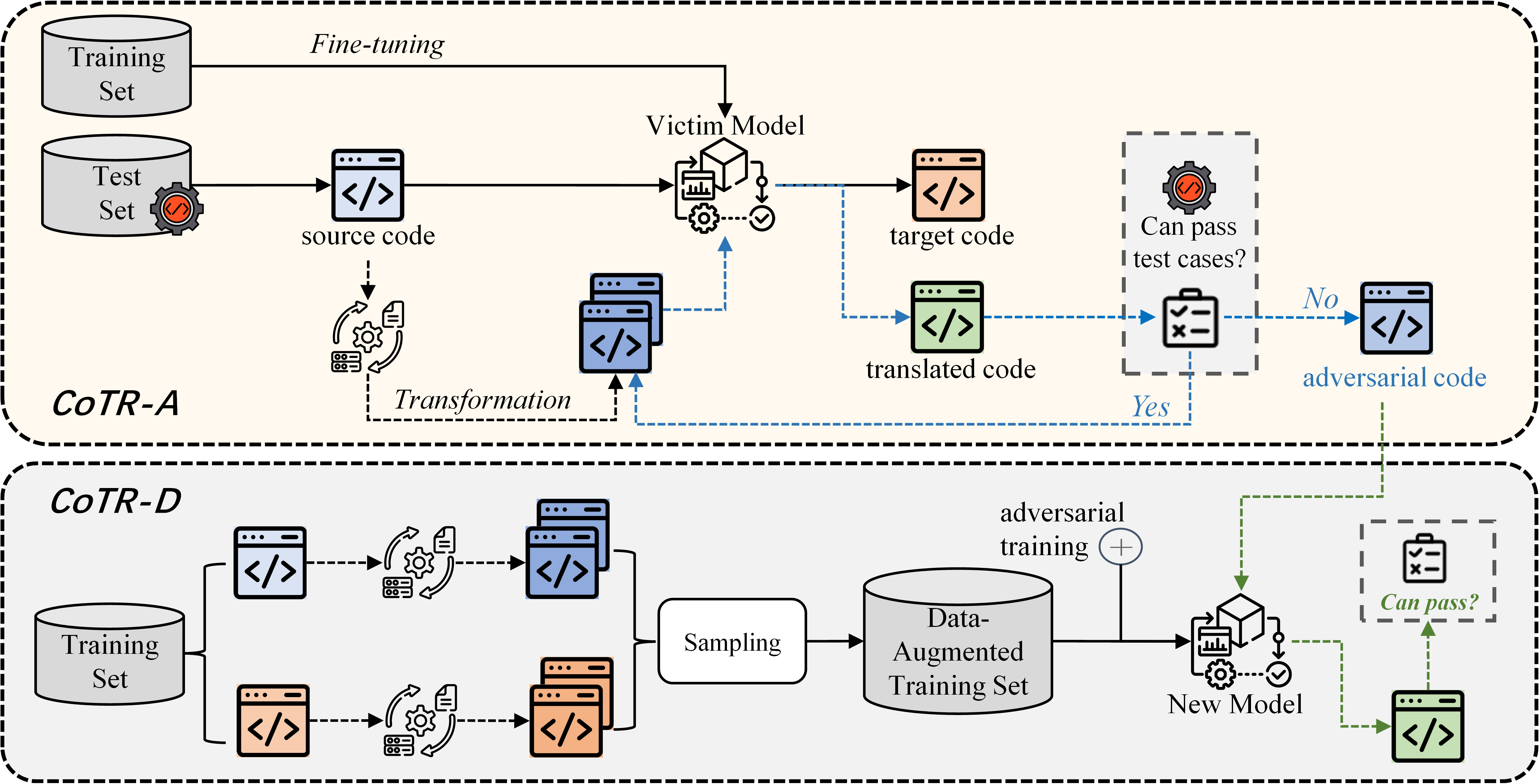}
\caption{The framework of {\tool}}
    \label{fig:model}
\end{figure*}

The framework of {\tool}  is illustrated in Figure~\ref{fig:model}, which consists of two major components, {\toolAttack} (the upper part of Figure~\ref{fig:model}) and {\toolDefense} (the lower part of Figure~\ref{fig:model}).  

\begin{figure}[h]
\centering
\includegraphics[width=0.4\textwidth]{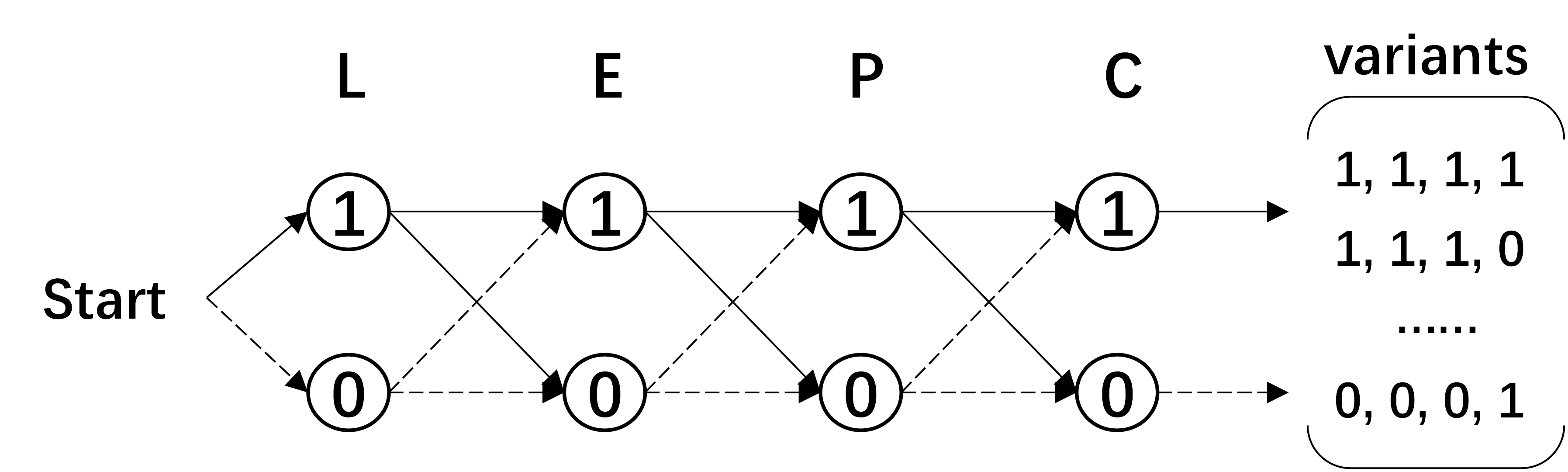}
\caption{Illustration of Candidate Code Snippet Generation}
    \label{fig:attack}
\end{figure}

\subsection{Attack Component: {\toolAttack}}

To assess the robustness of the pre-trained model, we first fine-tune it on a given dataset $\mathcal{D}_{train}$, resulting in the creation of the victim model $\mathcal{M}$. This model maps each source code $x$ to its corresponding target code $y = \mathcal{M}(x)$. Subsequently, we evaluate the performance of $\mathcal{M}$ on a designated test dataset $\mathcal{D}_{test}=\{(x_i, TC_{i})\}$ to determine the accuracy of $\mathcal{M}$ in translating the source code correctly.
To achieve this, we examine whether the translated target code $\mathcal{M}(x)$ for each $x$ from $\mathcal{D}_{test}$ successfully passes all the provided test cases ($TC$) for $x$. This evaluation process is formulated as follows.

\[
\boldsymbol{P}\left(\mathcal{M}, x_i, {TC}_i\right) = \left \{
\begin{aligned}
1, & \mbox{ If } \mathcal{M}(x_i) \mbox{ passes all test cases } {TC}_i,\\
0, & \mbox{ otherwise}.
\end{aligned}
\right.
\]

Intuitively, if the original output of $\mathcal{M}$ can successfully pass all the test cases, then the output code, even after experiencing minor perturbations to the input code, should also pass all the test cases. Therefore, in order to attack $\mathcal{M}$, we aim to generate an adversarial example $x_{adv}$ for a given input $x_i$, which should be sufficiently similar to $x_i$ but results in $\boldsymbol{P}\left(\mathcal{M}, x_{adv}, TC_i\right)=0$.

\begin{algorithm}[htbp]
\small
\caption{Adversarial Attack Algorithm} 
\label{alg:Framwork} 
\KwIn{
    Fine-tuned Code Translation Model $\mathcal{M}$;\\
    Code Translation DataSet with Test Cases $\mathcal{D}_{test}$;\\
    Transformation Rules $\mathcal{R}$;\\
    Transformation Constraint $\mathcal{G}$;\\
 }
\KwOut{
    Adversarial DataSet $\mathcal{D}_{adv}$;
}
Initialize Candidate Code Snippets $T \gets \emptyset$\;
Initialize Adversarial DataSet $\mathcal{D}_{adv}\gets \emptyset$\;
\For{each $\left(x, TC\right) \in \mathcal{D}_{test}$} {
    \If{$\boldsymbol{P}\left(\mathcal{M}, x, \textit{TC}\right)$ == 0}{
        $\mathcal{D}_{adv}\gets \mathcal{D}_{adv}\cup \{ x \}$\;
        break\;
    }
    $T \gets \mathcal{F}(x, \mathcal{R}, \mathcal{G})$\tcp{Generate candidate code snippets}
    \If{$T~is~\emptyset$}{
        $\mathcal{D}_{adv}\gets \mathcal{D}_{adv}\cup \{ x \}$\;
        break\;
    }
    $flag \gets$ 0\;
    \For{each $x_{t} \in T$} {
        \If{$\boldsymbol{P}\left(\mathcal{M}, x_{t}, \textit{TC}_x\right)$ == 0}{
            $\mathcal{D}_{adv}\gets \mathcal{D}_{adv}\cup \{ x_{t} \}$\;
            $flag \gets$ 1\;
            break\;
        }
    }
    \If{$flag$ == 0}{
        $\mathcal{D}_{adv}\gets \mathcal{D}_{adv}\cup \{ x \}$\;
    }
}
\Return $\mathcal{D}_{adv}$\; 
\end{algorithm}

Algorithm~\ref{alg:Framwork} provides the pseudo-code of {\toolAttack} to describe the detailed attack process. The initial step of {\toolAttack} is to generate all possible adversarial code snippets $T = \mathcal{F}(x, \mathcal{R}, \mathcal{G})$ for each sample in $\mathcal{D}_{test}$ through program transformation.
Subsequently, {\toolAttack} identifies the best code snippet for the original source code by minimizing the value of $\boldsymbol{P}$ to obtain the adversarial example. Formally, for each source code $x$ from $\mathcal{D}_{test}$,  we solve the following optimization problem
 \[
x_{adv}=\underset{\hat{x}\in T}{\arg\min } \boldsymbol{P}\left(\mathcal{M}, \hat{x}, TC_{x}\right)
\]
To obtain $x_{adv}$, we define $\mathcal{D}_{adv}= \{x_{adv} \mid x \in \mathcal{D}_{test}\}$. If an adversarial example $x_{adv}$ that successfully fools the model cannot be found, the original example $x$ is straightforwardly added to $\mathcal{D}_{adv}$.

\medskip
\noindent\textbf{Step 1. Generation of Candidate Code Snippets.}
As mentioned in Section~\ref{subsec:program}, for a given source code $x$, we construct its candidate code snippets using rule-based transformations. In the program analysis stage, we employ the third-party toolkit tree-sitter\footnote{\url{https://github.com/tree-sitter}}.
Regarding the transformation rules, we initially establish two constraints, denoted as $\mathcal{G}$:
(1) The variant code should maintain functional consistency, ensuring it passes all test cases as the original code does.
(2) The variant code should also be consistent with the original code in terms of informativeness and readability.

As presented in Table~\ref{tab:rules}, considering informativeness and readability, we exclusively generate candidates for the `Statement Exchange' category of transformation rules. It is worth noting that not all programming languages support `switch' statements (e.g., Python only introduced `match' statements as an alternative to `switch' statements in v3.10). Therefore, we consider the following four rules from the last category, `Statement Exchange', namely:

\begin{itemize}
    \item \textbf{Rule-L}: Loop Exchange; 
    \item \textbf{Rule-E}: Expression Exchange; 
    \item \textbf{Rule-P}: Permutation Exchange; 
    \item \textbf{Rule-C}: Condition Exchange. 
\end{itemize}

These rules distinguish themselves from mutation operators in mutation testing as they are designed to maintain the semantic, readability, and informativeness consistency between the variant code and the original code. Detailed functional descriptions and code examples for these rules are available in Table~\ref{tab:rules}.

Note that these transformation rules are not mutually exclusive;  after applying one rule, others can still be utilized to transform the source code. To improve the diversity of candidate code snippets, we take into account transformation sequences over \textbf{L}, \textbf{E}, \textbf{P}, \textbf{C}.
Each rule is capable of generating at most one code snippet. In instances where multiple locations in the code can be transformed (e.g., multiple occurrences of the ``+=" operator), we randomly select one for transformation.
A specific illustration is provided in Figure~\ref{fig:attack}, wherein the four rules serve as input parameters, each with a value of 0 or 1, indicating whether the rule is used for transformation or not. We exhaustively enumerate all strings over $\{L, E, P, C\}$, which gives a search space where each string denotes a transformation sequence. By applying these transformations, we can generate adversarial attacks.
  
It is essential to emphasize that the use of exhaustive heuristics aims to ensure the discovery of the most challenging adversarial examples. While heuristic algorithms can reduce search costs to some extent~\cite{yang2022natural}, their results may be influenced by prior assumptions and heuristic rules. Consequently, they may generate adversarial samples that are not optimal or most challenging.
In contrast, the exhaustive approach traverses all possible input variants, guaranteeing that no potential adversarial examples are overlooked. Although the computational complexity of this method is higher, it provides a more reliable assurance that the generated adversarial examples possess a high attack success rate.

\medskip
  \noindent\textbf{Step 2. Selection of Adversarial Example.}
This step is designed to identify the most effective adversarial examples within the search space, which can successfully deceive the victim model. As adversarial samples are typically generated from inputs that can be accurately processed by the victim model~\cite{zhang2022towards}, we exclude inputs that the victim model cannot process correctly.
For a given dataset $\mathcal{D}{test}$, we initially verify whether the code translated by the victim model $\mathcal{M}$ can pass all the test cases. If it fails to do so, we add this code to $\mathcal{D}_{adv}$ (Lines 4-6).

Next, {\toolAttack} generates all variant code snippets as candidates by exhaustively considering all possible combinations of the four transformation rules and verifying whether the generated candidates are empty. If the candidate set turns out to be empty, we include the original example in $\mathcal{D}_{adv}$ (Lines 7-10).
As we adopt a rule-based approach, not all code will be successfully transformed. Therefore, the time cost of using the search-based approach is deemed acceptable. For the generated candidates, we traverse through them to identify the adversarial example that can effectively attack the victim model (Lines 11-16).
Finally, in the event that no candidate can successfully attack the victim model, we add the original example to $\mathcal{D}_{adv}$ (Lines 17-18).


\subsection{Defense Component: {\toolDefense}}

As mentioned previously, it is crucial for adversarial code to retain functionality while maintaining the same level of informativeness and readability.
In order to augment the training dataset, we require source-target code snippet pairs where the program transformation rule is applied to both the source and target code.
%
%
To ensure the quality of the augmented data samples, constructing test cases for each sample becomes necessary. However, this process can be time-consuming, laborious, and even error-prone. Additionally, adding all variants to the augmentation dataset significantly increases the risk of model overfitting.

\medskip
\noindent\textbf{Data Augmentation.}
In light of this, we adopt a distinct data augmentation strategy in {\toolDefense}. Specifically, we employ a semantic distance-based sampling method to construct the augmented dataset more efficiently. To achieve this, we leverage the capabilities of CodeBERT~\cite{feng2020codebert} as a semantic feature extractor. This enables us to calculate the semantic distance between the original source code and its variants, as well as between the original target code and its variants. This process can be summarized as 
\[
\mathcal{D}_{aug} \sim 
\mathop{{\max}_{x^{\prime} \in \mathcal{F}(x, \mathcal{R}, \mathcal{G})\atop y^{\prime} \in \mathcal{F}(y, \mathcal{R}, \mathcal{G})}} 
\text{Distance}\left[f(x, x^{\prime}, y, y^{\prime})\right]
\]
where $f(x, x^{\prime}, y, y^{\prime}) = CodeBERT(x, x^{\prime}) + CodeBERT(y, y^{\prime})$.

We proceed to dataset augmentation $\mathcal{D}_{aug}$ by selecting the variant code snippet with the largest semantic distance from the $\mathcal{D}_{train}$. We calculate this distance by computing the cosine distance between the original source code and its variants, and also between the original target code and its variants. The sum of the two distance values is finally used.
By adopting this approach, we effectively enhance the diversity of the training set while mitigating the risk of overfitting.

\medskip
\noindent\textbf{Adversarial Training.}
It is observed in our empirical study (cf. Section~\ref{sect:rq3}) that data augmentation techniques have the potential to reduce the model's accuracy~\cite{bielik2020adversarial, zhou2022adversarial, yang2022important}.
To address this issue, we adopt a noisy-enhanced adversarial training method N-PGD based on Projected Gradient Descent~\cite{madrytowards}. 

In general, the PGD algorithm operates by iteratively perturbing the input data $x$ in the direction that maximizes the loss function, while ensuring that the perturbations remain within a specified epsilon bound. 
This iterative process is repeated for a fixed number of iterations, and the model parameters are updated during training based on the results.
The general principle of PGD can be summarized by
\[ 
\min_{\Theta_M} \mathbb{E}_{(x, y) \sim \mathcal{D}_{gra}}\left[\max _{\Delta x \in \Omega} \mathcal{L}\left(x+\Delta x, y ; \Theta_M\right)\right]
\]
where $\Delta x$ represents the perturbation applied to $x$, which is computed by the learning rate and the norm gradient of the $x$. $\Omega$ denotes the specified epsilon bound. The set $\mathcal{D}_{gra}$ refers to the gradient-based augmented dataset.

%% file: sections/4experiment.tex
\section{Experiments}
\label{sec:setup}

\subsection{Datasets}
To assess the effectiveness of our approach, we employ AVATAR, a compilation of the Java/Python dataset obtained from competitive programming sites, online platforms, and open-source repositories~\cite{ahmad2021avatar}.
In order to construct clean and high-quality datasets, as well as to facilitate the creation of test cases, we design four heuristic rules:

\begin{description}
\item[\textbf{H1}] Extract function-level code and perform syntax compilation check.
    
\item[\textbf{H2}] Remove code with input tokens such as `input()', `args *', etc.

   
\item[\textbf{H3}] Remove duplicate code.

\item[\textbf{H4}] Remove code with inconsistent method names for better readability.
\end{description}

After applying these heuristic rules, we obtain a set of 3,000 pairs of data samples, consisting of 2,600 pairs in the training set, 200 pairs in the validation set, and 200 pairs in the test set.
To ensure the quality of the test cases, we employ 10 postgraduate students, each has 3-5 years of programming experience. Each student is tasked to construct test cases for the samples in the test set, and each sample is evaluated using five test cases. To enhance the coverage of test cases, we implement a cross-checking process, wherein each student writes test cases for 20 code snippets and verifies their work with others. Additionally, students are permitted to search for relevant information and unfamiliar concepts on the Internet.
To prevent fatigue and maintain accuracy, we impose a limit on each student to write a maximum of 50 test cases within a half-day. Table~\ref{tab:code length} presents the statistical details of the Java and Python code in our dataset.

\begin{table}[htbp]
\label{tab:test}
	\centering
\caption{Length statistics of samples in the corpus}   
\begin{tabular}{lllllll}    
\toprule 
Language  & Avg. & Mode. & Median. & $<$128 & $<$256 \\    
\midrule  
Java & 100 & 79 & 90 & 72.5\% & 100\% \\
Python & 95 & 62 & 86 & 76.0\% & 100\% \\
\bottomrule   
\end{tabular}  
\label{tab:code length}
\end{table}

\subsection{Evaluation Metrics}

In this study, we consider different performance metrics to evaluate the code translation models, including

\begin{itemize}[leftmargin=*]
\item \textbf{BLEU}~\cite{papineni2002bleu}, which is widely used to measure the similarity between the translated and the reference code based on the $n$-gram precision.
    
\item \textbf{Code-BLEU}~\cite{ren2020codebleu}, an extension of BLEU  which considers keywords, syntax, and data flow of the translated code. 

\item \textbf{EM}, which measures the percentage of cases where the translated code exactly matches the reference code.
   
\item \textbf{Code-Exec}~\cite{liang2021lyra}, which examines the syntax of code to guarantee that there are no syntax errors, type errors, or other errors that could hinder the execution of the code. 

\item \textbf{P$_{s}$@1 (Pass@1)}~\cite{chen2021evaluating}, which is the percentage of the translated code that passes the test cases, i.e., the code which is deemed to be 
functionally correct.  
\end{itemize}

For the robustness of the model, we consider two specific metrics. 

\begin{itemize}[leftmargin=*]
\item \textbf{RP$_{s}$@1 (Robust Pass$_{s}$@1)}~\cite{wang2022recode}, which is  
the percentage of the translated code that passes the test cases after the adversarial attack. 
    
\item \textbf{RD$_{s}$@1 (Robust Drop$_{s}$@1)}~\cite{wang2022recode}, which means the relative performance change between P$_{s}$@1 and RP$_{s}$@1,  defined as
\[\text{RD$_{s}$@1}=1- \frac{\text{Robust Pass$_{s}$@1}}{\text{Pass@1}}\]
\end{itemize}

\subsection{Victim Pre-Trained Models}
We select ten widely used pre-trained models specialized for code translation tasks. These models can be classified into three groups based on their architecture: Encoder-only (Enc), Decoder-only (Dec), and Encoder-Decoder (Enc-Dec) models.
The Encoder-only models consist of CodeBERT~\cite{feng2020codebert}, GraphCodeBERT~\cite{guographcodebert}, and ContraBERT~\cite{liu2023contrabert}. The Decoder-only models encompass CodeGPT~\cite{lu1codexglue}, CodeGPT-adapter~\cite{lu1codexglue}, and CodeGen~\cite{nijkamp2022codegen}. Lastly, the Encoder-Decoder models include NatGen~\cite{chakraborty2022natgen}, CodeT5~\cite{wang2021codet5}, PLBART~\cite{ahmad2021unified}, and UniXcoder~\cite{guo2022unixcoder}. All pre-trained models and corresponding tokenizers are loaded from the official repository Huggingface.\footnote{\url{https://huggingface.co/models}} 
To ensure a fair comparison, we maintain consistent hyper-parameters for all models throughout our study. The hyper-parameters and their respective values are summarized in Table~\ref{Hyper-parameters}.

\begin{table}[h]
    \centering
    \caption{Hyperparameters and their value}
    \begin{tabular}{c|c||c|c}
    \toprule
       Hyperparameter  & Value &  Hyperparameter  & Value\\
     \midrule
      Optimizer & AdamW & Random Seed & 1,234 \\
      Learning Rate  & 5e-5 & Training batch size & 16 \\
      Beam size & 10 & Validation batch size & 16 \\
      Max input length & 350 & Max output length & 350 \\
      \bottomrule
    \end{tabular}
    \label{Hyper-parameters}
\end{table}

Our implementation is based on PyTorch 1.8, and the experiments are conducted on a machine with an Intel(R) Xeon(R) Silver 4210 CPU and the GeForce RTX 3090 GPU.


%% file: sections/5result.tex
\begin{table*}[]
 \caption{Comparison results between different PTMs}
 \begin{center}
 \setlength{\tabcolsep}{1mm}{
  \resizebox{0.96\textwidth}{!}{
\begin{tabular}{c|c|c|ccccc|ccccc}
  \toprule
\multirow{2}{*}{\textbf{Type}} & \multirow{2}{*}{\textbf{Model}} & \multirow{2}{*}{\textbf{Parameters}} & \multicolumn{5}{c|}{Java-to-Python} & \multicolumn{5}{c}{Python-to-Java}\\
& & & \textbf{$\mathit{BLEU}$} & \textbf{$\mathit{Code}$-$\mathit{BLEU}$} & \textbf{$\mathit{EM}$} & \textbf{$\mathit{Code}$-$\mathit{Exec}$} & \textbf{$\mathit{P_{s}@1}$} 
& \textbf{$\mathit{BLEU}$} & \textbf{$\mathit{Code}$-$\mathit{BLEU}$} & \textbf{$\mathit{EM}$} & \textbf{$\mathit{Code}$-$\mathit{Exec}$} & 
\textbf{$\mathit{P_{s}@1}$}\\
  \midrule
\multirow{4}{*}{Enc-Dec}
& NatGen & 223M & \textbf{84.36} & \textbf{80.57} & \underline{27.00} & \textbf{97.50} & \underline{73.50} & \underline{82.82} & \textbf{82.08} & \textbf{18.00} & \textbf{84.50} & \textbf{71.00} \\
& CodeT5 & 223M & \underline{83.30} & \underline{79.52} & \underline{27.00} & \textbf{97.50} & \textbf{76.00} & \textbf{83.11} & \underline{81.81} & 13.00 & \underline{84.00} & \underline{70.00} \\
& PLBART & 139M & 83.14 & 79.07 & 23.50 & 89.00 & 70.00 & 60.38 & 67.69 & 6.00 & 37.00 & 22.50 \\
& UniXcoder & 127M & 81.63 & 78.58 & 24.00 & 90.50 & 64.00 & 81.38 & 80.59 & 9.50 & 74.50 & 58.00 \\
\midrule
\multirow{3}{*}{Enc} 
&  CodeBERT & 173M & 75.12 & 71.99 & 10.50 & 66.00 & 40.50 & 76.03 & 74.87 & 5.00 & 45.00 & 29.50 \\
& GraphCodeBERT & 173M & 76.33 & 73.63 & 12.00 & 73.50 & 43.00 & 77.45 & 75.70 & 10.00 & 50.00 & 36.50 \\
& ContraBERT & 173M & 75.47 & 72.70 & 9.50 & 72.50 & 38.00 & 75.87 & 74.35 & 9.00 & 38.50 & 29.50 \\
\midrule
\multirow{3}{*}{Dec} 
& CodeGPT & 124M & 80.89 & 76.72 & 19.50 & 85.00 & 57.50 & 76.91 & 76.04 & \underline{17.00} & 67.00 & 49.50 \\
& CodeGPT-adapter & 124M & 82.18 & 78.20 & \textbf{27.50} & \underline{92.00} & 67.00 & 79.07 & 77.98 & 16.50 & 72.50 & 57.00 \\
& CodeGen & 355M & 81.35 & 78.09 & 17.00 & 90.50 & 59.50 & 79.50 & 79.03 & 15.00 & 68.50 & 51.50 \\
  \bottomrule
\end{tabular}}
 }
 \end{center}
 \label{tab:RQ1result}
\end{table*}

\section{Results} \label{sec:result}

\subsection{
\noindent\textbf{RQ1: How robust are existing pre-trained models under {\toolAttack}?}}


\subsubsection{Performance Comparison} 
Table~\ref{tab:RQ1result} presents the results, including evaluation metrics and model parameters, for comparative analysis. The best result is highlighted in bold, and the second-best result is underlined.
Our findings indicate that not all PTMs can effectively translate high-quality code. Specifically, NatGen and CodeT5 demonstrate significantly better performance compared to other models, achieving a pass@1 metric of more than 70; in contrast, CodeBERT and ContraBERT exhibit inferior performance, with pass@1 metrics of less than 40.

\noindent\textbf{Model architecture.} 
We observe that Encoder-Decoder models outperform Decoder-only and Encoder-only models, even though they have comparable numbers of parameters (124M$\sim$355M). This observation aligns with conclusions in the field of natural language generation.
 
\noindent\textbf{Evaluation metrics.} 
Our investigation reveals that the existing automatic evaluation metrics may not faithfully assess the functional correctness of translated code. For instance, the BLEU and CodeBLEU metrics of NatGen are higher than those of CodeT5, but the Pass@1 metric of its performance is lower.
This phenomenon is illustrated in Figure~\ref{fig:exam}, where translated code may resemble the reference code but fail compilation or some test cases. 





\begin{table}[]
 \caption{Comparison results between different attack methods}
 \begin{center}
 \setlength{\tabcolsep}{1mm}{
  \resizebox{0.48\textwidth}{!}{
\begin{tabular}{c|c|cc|cc}
  \toprule
\multirow{2}{*}{\textbf{Model}} & \multirow{2}{*}{\textbf{Attack}} & \multicolumn{2}{c|}{Java-to-Python} & \multicolumn{2}{c}{Python-to-Java}\\
& & \textbf{$\mathit{RP_{s}@1}$} & \textbf{$\mathit{RD_{s}@1}$} 
& \textbf{$\mathit{RP_{s}@1}$} & \textbf{$\mathit{RD_{s}@1}$} \\
  \midrule
\multirow{3}{*}{NatGen}
& RADAR & 70.50 & 4.08 & 68.50 & 3.52\\ 
& ALERT & 68.50 & 6.80 & 67.00 & 5.63\\ 
& CoTR-A & \textbf{59.50} & \textbf{19.05} & \textbf{60.00} & \textbf{15.49}\\
  \midrule
\multirow{3}{*}{CodeT5}
& RADAR & 71.00 & 6.58 & 62.50 & 9.29\\ 
& ALERT & 68.50 & 9.87 & 61.50 & 12.14\\ 
& CoTR-A & \textbf{60.00} & \textbf{21.05} & \textbf{60.00} & \textbf{14.29}\\
  \midrule
\multirow{3}{*}{PLBART}
& RADAR & 56.00 & 20.00 & \textbf{11.50} & \textbf{48.89}\\
& ALERT & 55.50 & 20.71 & 12.00 & 46.67\\
& CoTR-A & \textbf{47.00} & \textbf{32.86} & 17.00 & 24.44\\
  \midrule
\multirow{3}{*}{UniXcoder}
& RADAR & 56.00 & 12.50 & 46.50 & 19.83\\
& ALERT & 56.50 & 11.72 & 49.00 & 15.52\\
& CoTR-A & \textbf{52.50} & \textbf{17.97} & \textbf{40.50} & \textbf{30.17}\\
  \midrule
\multirow{3}{*}{CodeBERT}
& RADAR & 28.00 & 30.86 & 18.50 & 37.29\\
& ALERT & 29.50 & 27.16 & 22.00 & 25.42\\
& CoTR-A & \textbf{24.50} & \textbf{39.51} & \textbf{14.50} & \textbf{47.46}\\
  \midrule
\multirow{3}{*}{GraphCodeBERT}
& RADAR & \textbf{22.00} & \textbf{48.84} & \textbf{24.50} & \textbf{32.88}\\
& ALERT & 27.50 & 36.05 & 26.00 & 28.77\\
& CoTR-A & 24.50 & 43.02 & 26.00 & 28.77\\
  \midrule
\multirow{3}{*}{ContraBERT}
& RADAR & 24.00 & 36.84 & 20.50 & 30.44\\
& ALERT & 30.50 & 19.74 & 19.50 & 33.90\\
& CoTR-A & \textbf{22.50} & \textbf{40.79} & \textbf{18.50} & \textbf{37.29}\\
  \midrule
\multirow{3}{*}{CodeGPT}
& RADAR & 44.50 & 22.61 & 39.50 & 20.20\\
& ALERT & 44.50 & 22.61 & 38.00 & 23.23\\
& CoTR-A & \textbf{36.00} & \textbf{37.39} & \textbf{36.50} & \textbf{26.26}\\
  \midrule
\multirow{3}{*}{CodeGPT-adapter}
& RADAR & 62.50 & 6.72 & 51.50 & 9.65\\
& ALERT & 60.50 & 9.70 & 52.50 & 7.89\\
& CoTR-A & \textbf{45.50} & \textbf{32.09} & \textbf{40.00} & \textbf{29.82}\\
  \midrule
\multirow{3}{*}{CodeGen}
& RADAR & 54.50 & 8.40 & 47.50 & 7.77\\
& ALERT & 55.50 & 6.72 & 48.50 & 5.83\\
& CoTR-A & \textbf{45.00} & \textbf{24.37} & \textbf{37.50} & \textbf{27.18}\\
  \bottomrule
\end{tabular}}
 }
 \end{center}
 \label{tab:RQ2result}
\end{table}

\begin{table}[]
 \caption{Robustness Evaluation of LLMs}
 \begin{center}
 \setlength{\tabcolsep}{2mm}{
  \resizebox{0.45\textwidth}{!}{
\begin{tabular}{c|cccc}
  \toprule
\textbf{Task} & \textbf{Model} & \textbf{Pass@1} & \textbf{RP$_{s}$@1} & \textbf{RD$_{s}$@1}\\
  \midrule
\multirow{2}{*}{Java-to-Python}
& gpt-3.5-turbo & 87.50 & 80.50 & 8.00 \\
& CodeGeeX & 36.50 & 15.00 & 58.90 \\
\midrule
\multirow{2}{*}{Python-to-Java}
& gpt-3.5-turbo & 79.50 & 56.00 & 29.56 \\
& CodeGeeX & 30.50 & 19.00 & 37.70 \\
  \bottomrule
\end{tabular}}
 }
 \end{center}
 \label{llm}
\end{table}


\subsubsection{Robustness Comparison} 

{\toolAttack} is a syntactic transformation-based attack method that satisfies the constraints described in Section~\ref{sec:method}. To compare its effectiveness with other token-based attack methods, we select RADAR~\cite{yang2022important} and ALERT~\cite{yang2022natural} as baselines. ALERT utilizes CodeBERT and GraphCodeBERT to generate natural candidates and employs a combination of greedy search and genetic algorithm for optimization. RADAR considers semantic equivalence, typos, and visual similarity, as simple typos are known to be significant in code refactoring.
It is worth noting that {\toolAttack} is compatible with existing token-based methods and can be combined with them to provide a more comprehensive evaluation of PTMs' robustness. In our empirical study (Table~\ref{tab:RQ2result}), we present metrics (such as RP${s}$@1 and RD${s}$@1). Higher RP${s}$@1 values or lower RD${s}$@1 values indicate greater model robustness. The best results are highlighted in \textbf{boldface}. Furthermore, as the size of language models and training data continue to grow, large language models (LLMs) demonstrate various emergent behaviors~\cite{weiemergent} (here LLMs refer to models with 10B+ parameters). One such ability is zero-shot learning, which allows models to answer within a specific instruction or prompt~\cite{ouyang2022training}. In particular, LLMs have achieved excellent performance and demonstrated great potential on code translation tasks~\cite{chen2021evaluating}.
To further verify the robustness of LLMs and the effectiveness of our attack method, we discuss the zero-shot performance of CodeGeeX~\cite{zheng2023codegeex}\footnote{\url{https://codegeex.cn/codeTranslator}} and ChatGPT (gpt-3.5-turbo\footnote{\url{https://platform.openai.com/docs/models/gpt-3-5}}) in Table~\ref{llm}.

\noindent\textbf{Robustness degradation.} 
We evaluate model robustness using the RD${s}$@1 metric, where a higher RD${s}$@1 value indicates lower robustness. For example, the CodeT5 model achieves a P${s}$@1 value of 76\% when translating Java to Python (cf. Table~\ref{tab:RQ1result}). However, under the {\toolAttack} attack, RP${s}$@1 decreases to 60\%, and the RD$_{s}$@1 increases to 21.05\%, indicating a significant 21.05\% performance reduction.
Comparing Table~\ref{tab:RQ1result} and Table~\ref{tab:RQ2result}, we observe performance degradation across all PTMs, with RD${s}$@1 values ranging from 14.29\% to 47.46\%. We conclude that these models are generally \textit{not} robust for code translation. Among the different models, NatGen and CodeT5 exhibit the best robustness performance (their RD${s}$@1 value can also be maintained at around 14.29\% to 21.05\% under {\toolAttack}'s attack). Conversely, the Encoder-only models show the least robustness.
Furthermore, we find that ChatGPT performs well in zero-shot scenarios and outperforms NatGen and CodeT5. However, it still exhibits robustness issues, as seen in both CodeGeeX and ChatGPT.


\noindent\textbf{Attack effectiveness.} Table~\ref{tab:RQ2result} shows that {\toolAttack} generally outperforms RADAR and ALERT in terms of attack effectiveness, except for PLBART and GraphCodeBERT.
The vulnerability of the Encoder-only model to token-based attacks is noteworthy, likely due to the random initialization of its decoder parameters and the lack of pre-training.
Additionally, Shi et al.~\cite{shi2023towards} suggest that lower model layers tend to concentrate on lexical properties, while higher layers focus on syntactic and semantic properties. This finding may explain the superior performance of syntax-based attacks compared to token-based attacks.

\noindent\textbf{Pre-training techniques.}
We also evaluate the impact of different pre-training techniques on model robustness. NatGen incorporates a de-naturalizing pre-training task, focusing on the naturalness of code, which leads to performance that outperforms CodeT5. ContraBERT incorporates contrast learning, leading to better RD@1 results but worse RP$_{s}$@1 results compared to GraphCodeBERT. CodeGPT-adapter utilizes adapter learning and demonstrates improved performance over CodeGPT.
From Table~\ref{tab:RQ2result}, we observe that these pre-training techniques may be effective in defending against token-based attacks but are less effective against syntax-based attacks like {\toolAttack}.
 
\subsubsection{Human Study} 
To evaluate the quality of adversarial code, we further conduct a human evaluation study. We collect code snippets that can be attacked by RADAR, ALERT, and {\toolAttack} in the above experiments, resulting in a total of 159 pairs. 
We invite five graduated students who have 3$\sim$5 years of experience in Java and Python to participate in the evaluation.
\begin{figure}[h]
	\centering
	\includegraphics[width=0.48\textwidth]{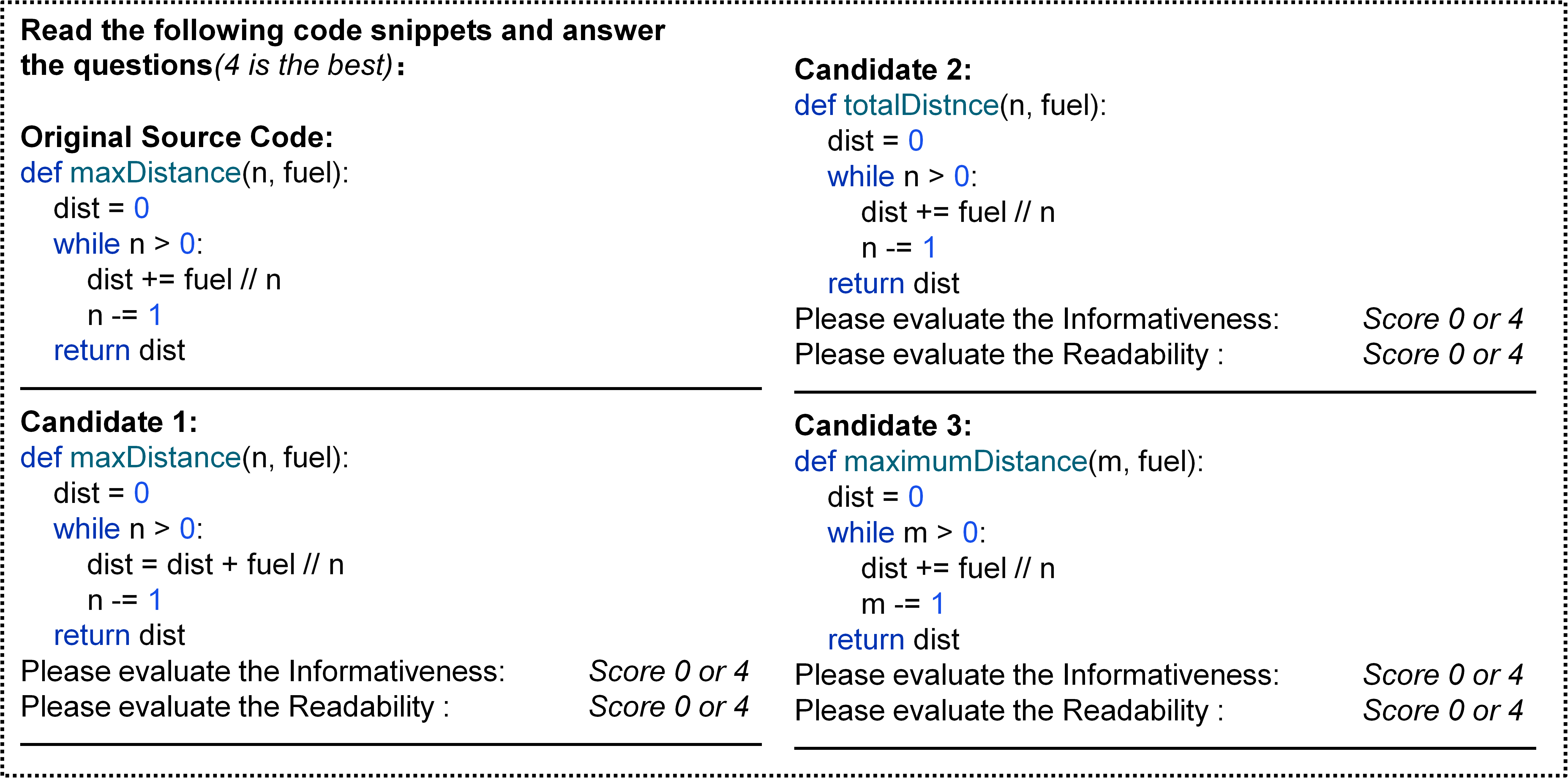}
	\caption{Sample  questionnaire  used in the  human evaluation}
	\label{fig:human}
\end{figure}
To conduct the evaluation, we generate a questionnaire (shown in Figure~\ref{fig:human}) for each code snippet and ask each participant to score the informativeness and readability of three adversarial examples generated by RADAR, ALERT, and {\toolAttack}. The scores range from 0 to 4, with higher scores indicating better quality. To ensure a fair comparison, the source of the adversarial code is hidden from the participants, and the order of the questionnaires is randomized. The workload of each participant is restricted, not exceeding 50 code snippets in half a day, to ensure the quality of evaluation.

\begin{table}[h]
	\begin{center}
		\setlength{\tabcolsep}{4mm}{
			\caption{Results of our human study}
			\label{tab:human}
			\begin{tabular}{cccc}
				\toprule
				\textbf{Approach} & \textbf{Informativeness} & \textbf{Readability} \\
				\midrule
				RADAR  & 3.25   &  3.47 \\
				ALERT  & 3.45   &  3.50 \\
				{\toolAttack} & \textbf{3.64}  & \textbf{3.55}   \\
				\bottomrule
			\end{tabular}
		}
	\end{center}
\end{table}

The results of the human evaluation study are presented in Table~\ref{tab:human}, which shows the average scores given by the participants for each adversarial code snippet in terms of informativeness and readability. We observe that {\toolAttack} outperforms RADAR and ALERT on both aspects, with an improvement of 0.39 and 0.08 respectively. This suggests that {\toolAttack} generates adversarial code that is more informative and readable than those generated by RADAR and ALERT, further verifying the superiority of our approach.

\begin{tcolorbox}[width=1.0\linewidth, title={Summary of RQ1}]
(1) The empirical study reveals that the existing PTMs are generally not robust for code translation tasks.
(2) The syntactic transformation-based attack method  {\toolAttack} can outperform token-based attacking methods on most models. 
\end{tcolorbox}







\begin{table*}[h]
 \caption{Comparison results  between different defense methods in the Java-to-Python dataset}
 \begin{center}
  \vspace{-1mm}
 \setlength{\tabcolsep}{2mm}{
  \resizebox{1.0\textwidth}{!}{
\begin{tabular}{c|c|cccc|cccc|cccc}
  \toprule
\multirow{2}{*}{\textbf{Type}} & \multirow{2}{*}{\textbf{Model}} & \multicolumn{4}{c|}{\textbf{$\mathit{P_{s}@1}$}} & \multicolumn{4}{c|}{\textbf{$\mathit{RP_{s}@1}$}} & \multicolumn{4}{c}{\textbf{$\mathit{RD_{s}@1}$}}\\
 &  & Original & DA & AT & \toolDefense & Original & DA & AT & \toolDefense & Original & DA & AT & \toolDefense\\
  \midrule
\multirow{4}{*}{Enc-Dec}
& NatGen & 73.50 & 74.00 & \textbf{80.50} & \underline{79.50} & 59.50 & \underline{70.50} & 69.50 & \textbf{75.00} & 19.50 & \textbf{4.73} & 14.29 & \underline{5.66} \\
& CodeT5 & \underline{76.00} & 69.50 & \textbf{77.50} & 74.50 & 60.00 & \underline{69.00} & 66.50 & \textbf{74.00} & 21.05 & \underline{0.72} & 14.19 & \textbf{0.67} \\
& PLBART & \textbf{70.00} & 51.50 & \underline{69.50} & 68.00 & 47.00 & 50.00 & \underline{56.00} & \textbf{65.50} & 32.86 & \textbf{2.91} & 19.42 & \underline{3.68} \\
& UniXocder & 64.00 & \underline{69.00} & \textbf{71.50} & \underline{69.00} & 52.50 & \underline{67.50} & 66.00 & \textbf{68.50} & 17.97 & \underline{2.17} & 7.69 & \textbf{0.72} \\
\midrule
\multirow{3}{*}{Enc}
& CodeBERT & 40.50 & 43.00 & \textbf{47.50} & \underline{43.50} & 24.50 & \underline{39.00} & \textbf{40.00} & \textbf{40.00} & 39.51 & \underline{9.30} & 15.79 & \textbf{8.75} \\
& GraphCodeBERT & 43.00 & 41.50 & \textbf{48.50} & \underline{44.50} & 24.50 & 36.50 & \underline{39.00} & \textbf{40.50} & 43.02 & \underline{12.05} & 19.59 & \textbf{8.99} \\
& ContraBERT & 38.00 & 42.50 & \underline{46.00} & \textbf{48.50} & 22.50 & \underline{40.00} & 35.00 & \textbf{42.00} & 40.79 & \textbf{5.88} & 23.91 & \underline{15.48} \\
\midrule
\multirow{3}{*}{Dec}
& CodeGPT & 57.50 & \underline{58.00} & 53.50 & \textbf{61.50} & 36.00 & \underline{53.50} & 41.50 & \textbf{55.50} & 37.39 & \textbf{7.76} & 22.43 & \underline{9.76} \\
& CodeGPT-adapter & \underline{67.00} & 63.50 & 61.50 & \textbf{67.50} & 45.50 & \underline{61.00} & 46.00 & \textbf{64.00} & 32.09 & \textbf{3.94} & 25.20 & \underline{5.19} \\
& CodeGen & 59.50 & \textbf{67.00} & \underline{61.00} & \textbf{67.00} & 45.00 & \textbf{66.00} & \underline{57.00} & \textbf{66.00} & 24.37 & \textbf{1.49} & \underline{6.56} & \underline{1.49} \\
  \bottomrule
\end{tabular}}
 }
 \end{center}
 \label{tab:RQ4result-1}
\end{table*}
\begin{table*}[h]
 \caption{Comparison results between different defense methods in the Python-to-Java dataset}
 \begin{center}
\vspace{-1mm}
 \setlength{\tabcolsep}{2mm}{
  \resizebox{1.0\textwidth}{!}{
\begin{tabular}{c|c|cccc|cccc|cccc}
  \toprule
\multirow{2}{*}{\textbf{Type}} & \multirow{2}{*}{\textbf{Model}} & \multicolumn{4}{c|}{\textbf{$\mathit{P_{s}@1}$}} & \multicolumn{4}{c|}{\textbf{$\mathit{RP_{s}@1}$}} & \multicolumn{4}{c}{\textbf{$\mathit{RD_{s}@1}$}}\\
 &  & Original & DA & AT & \toolDefense & Original & DA & AT & \toolDefense & Original & DA & AT & \toolDefense \\
  \midrule
\multirow{4}{*}{Enc-Dec}
& NatGen & 71.00 & \textbf{73.00} & \underline{72.00} & \textbf{73.00} & 60.00 & 66.50 & \textbf{68.00} & \underline{67.50} & 15.49 & 8.90 & \textbf{5.56} & \underline{7.53} \\
& CodeT5 & 70.00 & 61.50 & \textbf{72.50} & \underline{71.00} & 60.00 & 58.00 & \textbf{67.00} & \underline{66.50} & 14.29 & \textbf{5.69} & 7.59 & \underline{6.34} \\
& PLBART & 22.50 & \underline{44.00} & 16.50 & \textbf{56.50} & 17.00 & \underline{41.50} & 15.50 & \textbf{55.00} & 24.44 & \underline{5.68} & 6.06 & \textbf{2.65} \\
& UniXocder & 58.00 & \underline{63.50} & 55.00 & \textbf{66.50} & 40.50 & \underline{56.50} & 47.50 & \textbf{58.00} & 30.17 & \textbf{11.02} & 13.64 & \underline{12.78} \\
\midrule
\multirow{3}{*}{Enc}
& CodeBERT & 29.50 & \underline{32.00} & \underline{32.00} & \textbf{36.00} & 15.50 & \underline{31.00} & 27.00 & \textbf{33.00} & 47.46 & \textbf{3.13} & 15.63 & \underline{8.33} \\
& GraphCodeBERT & 36.50 & \textbf{43.50} & 33.00 & \underline{41.00} & 26.00 & \underline{39.50} & 29.00 & \textbf{40.00} & 28.77 & \underline{9.20} & 12.12 & \textbf{2.44} \\
& ContraBERT & 29.50 & \textbf{38.50} & 36.00 & \underline{36.50} & 18.50 & \textbf{38.50} & 33.50 & \underline{36.00} & 37.29 & \textbf{0.00} & 6.94 & \underline{1.37} \\
\midrule
\multirow{3}{*}{Dec}
& CodeGPT & \underline{49.50} & 47.50 & \underline{49.50} & \textbf{50.00} & 36.50 & 43.00 & \underline{44.50} & \textbf{45.00} & 26.26 & \textbf{9.47} & 10.10 & \underline{10.00} \\
& CodeGPT-adapter & \textbf{57.00} & 51.50 & \underline{55.00} & \textbf{57.00} & 40.00 & 47.50 & \underline{50.50} & \textbf{51.00} & 29.82 & \textbf{7.77} & \underline{8.18} & 10.53 \\
& CodeGen & 51.50 & 54.00 & \underline{56.50} & \textbf{60.50} & 37.50 & \underline{50.50} & 49.00 & \textbf{57.00} & 27.18 & \underline{6.48} & 13.27 & \textbf{5.79} \\
  \bottomrule
\end{tabular}}}
 \end{center}
 \label{tab:RQ4result-2}
\end{table*}


\subsection{
\noindent\textbf{RQ2: How effective is {\toolDefense} in improving the robustness of existing PTMs for code translation?}
} \label{sect:rq3}


To assess the impact of {\toolDefense} on enhancing the robustness of PTMs for code translation, we conducted a comparison for different models in terms of the Pass@1, RP${s}$@1, and RD${s}$@1 metrics. These models include the original model without any defense mechanism, the model with data augmentation, the model with adversarial training, and the model with our proposed {\toolDefense} method. The experimental results for the Python to Java translation task can be found in Table~\ref{tab:RQ4result-1}, while the results for the Java to Python translation task are presented in Table~\ref{tab:RQ4result-2}.

 
\noindent\textbf{Data augmentation (DA).} 
DA has demonstrated effectiveness in enhancing model robustness; however, it may not be adequate to guarantee optimal performance. As observed in the results, DA can provide an advantage in terms of the RD${s}$@1 metric, indicating improved robustness. However, there could be a potential degradation in the P${s}$@1 metric, signifying reduced performance. This phenomenon can be attributed to certain models excessively emphasizing the augmented data during the fine-tuning process, which can be understood from the perspective of data distribution.

\begin{figure}[h]
	\vspace{-0.5em}
	\centering
	\subfigure[Dimensional distribution on the Java dataset]{%
		\includegraphics[width=0.22\textwidth]{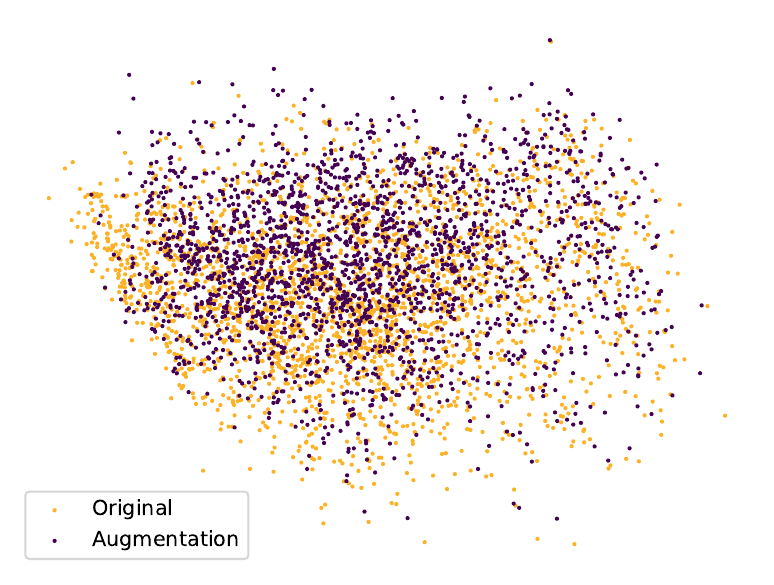}%
	}%
	\hfill
	\subfigure[Dimensional distribution on the Python dataset]{%
		\includegraphics[width=0.22\textwidth]{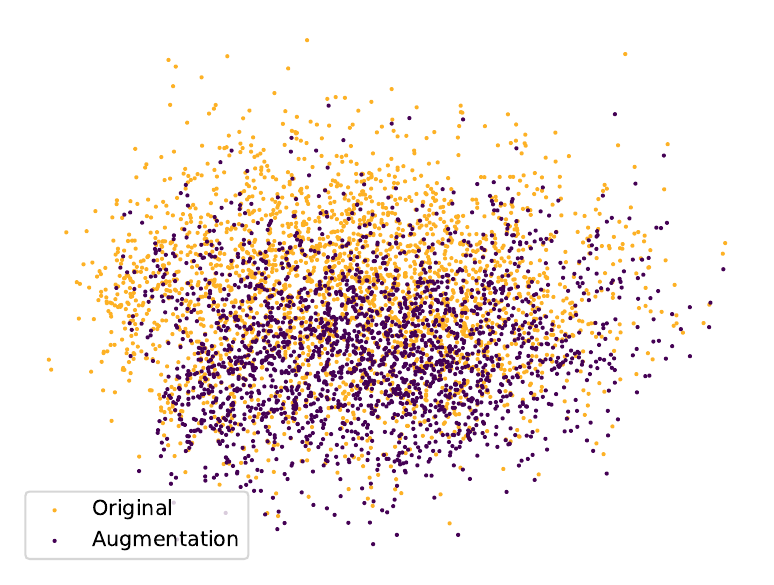}%
	}%
	\caption{Dimensional distribution of original and augmentation data}
	\vspace{-0.5em}
	\label{fig:distribution}
\end{figure}

DA is effective in improving model robustness, but it may not be sufficient to ensure its performance. As seen in the results, DA is able to show an advantage in the RD$_{s}$@1 metric, but there may be a performance degradation in the P$_{s}$@1 metric. The reason is that some of the models are overly about the augmented data part in the fine-tuning process, which can be explained from the perspective of data distribution. 
Indeed we analyze the distribution relationship between the original and augmented data by visualizing the semantic feature representations using CodeBERT. 
We apply PCA~\cite{mackiewicz1993principal} to obtain graphs of the different datasets. 
We map each sample into a 512-dimension vector through CodeBERT and the mean-pooling operation, and then project the vector into a two-dimensional plane using PCA, as shown in Figure~\ref{fig:distribution}.

The results of the data distribution analysis are shown in Figure ~\ref{fig:distribution}. 
From the distributions of the original and augmented datasets,  
We can see that they are very similar. The slight difference lies in, for instance, for the distribution of the python dataset, the original data is  skew towards the upper part of the semantic space, while the enhanced data is  skew towards the lower part. 
This indicates that the data enhancement method successfully maintains the original data distribution and expands it accordingly.

\smallskip
\noindent\textbf{Adversarial training (AT).} 
AT has demonstrated effectiveness in enhancing model performance; however, it is often less robust than data augmentation (DA) in terms of improving model robustness. As observed in the results, AT can provide an advantage in the P${s}$@1 metric, indicating improved performance. However, there could be a potential degradation in the RD${s}$@1 metric, signifying reduced robustness. This finding suggests that gradient-based AT is useful in improving the model's performance, but it may not be sufficient on its own.


Our proposed {\toolDefense} approach combines the strengths of both DA and AT techniques to improve the robustness and performance of pre-trained models (PTMs) in code translation tasks. By leveraging DA to generate diverse training data and AT to train models on adversarial examples, {\toolDefense} ensures that the model's performance remains stable while significantly enhancing its robustness against various types of attacks. For instance, the RD${s}$@1 values for most of the models are kept within 10\%, while their P${s}$@1 and RP$_{s}$@1 values are better than the original models.

\begin{tcolorbox}[width=1.0\linewidth, title={Summary of RQ2}]
Employing data augmentation or adversarial training techniques alone may damage the model's performance or robustness. However, our proposed {\toolDefense} approach effectively combines these two techniques, resulting in improved model robustness without sacrificing the translation accuracy. 
 \end{tcolorbox}

%% file: sections/6discussion.tex
\section{Threats to validity}
\label{sec:discussion}

\noindent\textbf{Threats to internal validity.} 
Firstly, we mitigate implementation errors by conducting thorough checks on our implementation and utilizing mature libraries. Additionally, we have ensured the functionality of the variant code generated by {\toolAttack} by test cases.
Secondly, to ensure a comprehensive evaluation of different model types, we have chosen ten state-of-the-art models by covering three diverse types of models. 

\noindent\textbf{Threats to external validity.} 
Our dataset is derived from code competitions, and thus may not fully reflect the complexity of real-world scenarios. However, it provides valuable initial insights into the challenges of robust code translation. Importantly, our approach is language-independent, and the proposed enhancements can be applicable to different programming languages.
In future research, we plan to expand our study to include more diverse and complex programs to validate the effectiveness of the proposed enhancements on a larger scale.

\noindent\textbf{Threats to construct validity.} 
Performance measure selection is the main construct threat. To mitigate this, we selected five widely used performance measures to evaluate the translation quality of our models. Additionally, to assess the robustness of our models against adversarial attacks, we introduce two specific metrics, RP${s}$@1 and RD${s}$@1, which focus on the success rate and diversity of the attacks, respectively. Furthermore, we conducted a human study to analyze the quality of our generated adversarial code.


%% file: sections/7related.tex
\section{Related Work}
\label{sec:related}



\subsection{Code Translation}

Early studies utilized rule templates or statistical methods to perform translations between different programming languages. For instance, phrase-based models were employed to translate code from C\# to Java or from Python2 to Python3~\cite{nguyen2013lexical, karaivanov2014phrase}.
In a later study, An \textit{et al.}~\cite{an2018automatic} proposed a rule-based approach that inferred syntactic transformation rules and API mappings to automatically translate Java code to Swift. 
Zhong et al.~\cite{zhong2010mining} explored the use of Application Programming Interfaces (APIs) in the context of code translation.
However, these approaches are typically limited to a few specific language pairs and often require the creation of parallel datasets either manually or through rule-based tools. 

In recent years, attention has shifted towards neural network based approaches (in particular, pre-trained models) for code translation. 
Roziere et al.~\cite{roziere2020unsupervised} proposed TransCoder, an unsupervised pre-trained model based on unsupervised machine translation. 
Roziere et al.~\cite{lachaux2021dobf} showed that augmenting TransCoder with de-obfuscated targets can significantly improve performance. 
Liu et al.~\cite{liu2023syntax} proposed SDA-Trans, a syntax and domain-aware model for program translation. 
Meanwhile, supervised approaches have also proven successful, and the ten code pre-training models mentioned in this paper have all achieved impressive results when used for code translation as a downstream task after fine-tuning.

\subsection{Adversarial Attack and Defense on Code-related Models}
The robustness of neural network models has been extensively studied, particularly in image classification tasks. However, there is also a growing body of research focusing on code-related tasks, such as source code classification (Code$\mapsto$Label)~\cite{tian2021generating}, code summarization (Code$\mapsto$NL)~\cite{zhou2022adversarial}, and code generation task (NL$\mapsto$Code)~\cite{yang2022important, wang2022recode}.

Adversarial attacks on code can manifest in two forms: token-based attacks and syntax-based attacks. 
Token-based attacks predominantly focus on code identifiers and manipulate the model by replacing tokens with equivalent semantics.
For instance, 
Zhang et al.~\cite{zhang2020generating} proposed MHM, which utilizes Metropolis-Hastings sampling-based identifier renaming.  
Zeng et al.~\cite{zeng2022extensive} employed a wide range of NLP-based adversarial attack methods to evaluate pre-trained models and discovered that random attack methods can outperform carefully designed adversarial attack methods in most cases.
Recent research has increasingly emphasized addressing the naturalness aspect of adversarial examples. 
Yang et al.~\cite{yang2022natural} proposed a naturalness-aware attack called ALERT, which generates multiple natural candidates using GraphCodeBERT and CodeBERT.
Zhou et al.~\cite{zhou2022adversarial} proposed ACCENT, which generates multiple natural candidates using the word2vec.
Zhang et al.~\cite{zhang2022towards} introduced CARROT, an optimization-based attack technique that assesses and improves the robustness of deep program processing models.
Yang et al.~\cite{yang2022important} proposed RADAR, which generates semantic and visual similar adversarial examples for code generation. 
Jha and Reddy~\cite{jha2023codeattack} proposed CodeAttack, which finds the most vulnerable tokens and then substitutes these vulnerable tokens to generate adversarial examples.

Syntax-based attacks are primarily concerned with the syntax of the code and manipulate the model through transformations that preserve syntactic equivalence.
Pour et al.~\cite{pour2021search} introduced a search-based testing framework for deep neural networks of source code embedding. Their framework focused on "for-loop enhance" and "if-loop enhance" to target code syntax. They applied this framework to tasks such as method name prediction, code captioning, code search, and code documentation generation.
Rabin et al.~\cite{rabin2021generalizability} conducted an evaluation of multiple syntactic transformations on code search, code summarization, and code analogies. However, their study did not consider combinations of these transformations.


Adversarial defense on code models can be categorized as either active or passive. Active defense approaches involve re-training models with adversarial examples to enhance their robustness. For instance, Zhang et al.~\cite{zhang2020generating} proposed adversarial training as an active defense method for code translation tasks.
In contrast, passive defense approaches aim to restore model performance without re-training or modifying the model. Zhou et al.~\cite{zhou2022adversarial} introduced a lightweight adversarial training method called the mask training algorithm. 
Yang et al.~\cite{yang2022important} also proposed a passive defense approach for code generation tasks through method name generation.

In contrast to previous work, we focus on program transformation based attacks instead of token-based attacks. Additionally, we investigate the impact of combining different program transformation methods, providing insights into the factors that contribute to the non-robustness of existing pre-trained models.
Furthermore, we explore and employ a variety of defensive approaches to enhance model robustness and generalization in the face of adversarial attacks. Our study aims to contribute to a comprehensive understanding of the vulnerabilities and defenses in the context of code translation tasks.

%% file: sections/8conclusion.tex
\section{Conclusion}
\label{sec:conclusion}

In this study, we have conducted a thorough investigation of the robustness of pre-trained models (PTMs) in code translation tasks. We present {\tool}, a novel approach that aims to assess and enhance the robustness of these models. Our research exposes the limitations of existing PTMs, including large language models (LLMs) such as CodeGeeX and ChatGPT 3.5, in effectively handling code translation tasks.
To address these limitations, we propose {\toolDefense}, a defense mechanism that demonstrates promising results in improving the robustness and generalization of PTMs. Our findings provide valuable insights into the challenges and potential solutions for building more robust code translation models.

In future work, we plan to develop a more robust pre-trained model that can handle different programming styles and syntax conventions. We also plan to explore the use of other techniques, such as program repair or LLMs, to improve the effectiveness and robustness of pre-trained models in handling code translation tasks.